The NASA/IPAC Teacher Archive Research Program (NITARP)

L. M. Rebull (Caltech-IPAC/IRSA and NITARP), M. Fitzgerald (Edith Cowan Institute for Education Research), T. Roberts (Caltech-IPAC/ICE), D. A. French (Wilkes University), W. Laurence (Creato-osity), V. Gorjian (Jet Propulsion Laboratory and NITARP), G. K. Squires (Caltech-IPAC/ICE)

**Abstract**

NITARP, the NASA/IPAC Teacher Archive Research Program, partners small groups of predominantly high school educators with research astronomers for a year-long research project. This paper presents a summary of how NITARP works and the lessons learned over the last 13 years. The program lasts a calendar year, January to January, and involves three ~week-long trips: to the American Astronomical Society (AAS) winter meeting, to Caltech in the summer (with students), and back to a winter AAS meeting (with students) to present their results. Because NITARP has been running since 2009, and its predecessor ran from 2005-2008, there have been many lessons learned over the last 13 years that have informed the development of the program. The most critical is that scientists must see their work with the educators on their team as a partnership of equals who have specialized in different professions. NITARP teams appear to function most efficiently with approximately 5 people: a mentor astronomer, a mentor teacher (who has been through the program before), and 3 new educators. Educators are asked to step into the role of learner and develop their question-asking skills as they work to develop an understanding of a subject in which they will not have command of all the information and processes needed. Critical to the success of each team is the development of communication skills and fluid plan of action to keep the lines of communication open. This program has allowed more than 100 educators to present more than 60 total science posters at the AAS.

**Introduction**

NITARP, the NASA/IPAC Teacher Archive Research Program[1], fosters partnerships between teachers and research astronomers. Small groups of educators from all over the United States are paired with a professional astronomer for a year-long original research project. NITARP works with teachers specifically because of the influence possible through them to their students and communities. Most of the educators are high school classroom educators, though some middle school and informal educators have participated. NITARP's goals are to provide a professional development experience for teachers that enables them to experience the real research process, through which their understanding of the nature of research is deepened and ultimately their current and future students are affected via changes in teaching styles.

Participating in an authentic scientific research project as a high school student may help keep students in the leaky STEM pipeline. Historically, students have felt unable to do science; however, current research shows that while high- and middle-school students feel they are capable of doing science, they choose not to do science (Kitts 2009).  At the same time, high school students are taking more math and science classes than in previous years; they are also earning higher grades (Hill, Corbet, & St. Rose 2010).  This is particularly true for women and underrepresented minorities (Hill, Corbett, & St. Rose 2010).  Those women and underrepresented students, however, do not go on to STEM majors in college (Hill, Corbett, & St. Rose 2010). What makes a STEM student stay in the field? Over half (53%) of all college STEM majors conducted a research project while an undergraduate student (Russel 2006).  Students who have participated in a research project at an undergraduate level are more likely to stay in a STEM field (Russel, Hancock, & McCullough 2007), though it is admittedly unclear if this can be extended to high school students.  If a research project at an undergraduate level can retain STEM students, perhaps more authentic science experiences at the high school level (such as NITARP) can retain STEM students at an earlier stage.

---

[1] http://nitarp.ipac.caltech.edu

Teachers must be ready to support authentic science experiences in pre-college settings. As many as two-thirds of science educators do not have graduate or undergraduate degrees in science (Schools and Staffing Survey 2012 as cited in Marder 2017); even those who hold undergraduate degrees in science are unlikely to have participated in authentic scientific research (National Academies of Science, Engineering, and Medicine 2015; National Research Council 2006, 2012).  Especially in the context of the Next Generation Science Standards (NGSS 2013) and other reform efforts focusing on incorporation of not only more inquiry-based activities but also authentic science experiences in the classroom, this means that teachers are being asked to teach something they have not experienced themselves (see, e.g., Crawford 2014).  Professional development opportunities that expose educators to authentic science, such as NITARP, are sorely needed.

NITARP's predecessor, the Spitzer Space Telescope Research Program for Teachers and Students, ran from 2004-2008. This program granted small amounts of Director's Discretionary Time (DDT) on the Spitzer Space Telescope (Werner et al. 2004) to teacher teams; they did scientific research using these new Spitzer data.  Leveraging on a well-established teacher professional development program, the Spitzer opportunity was offered to graduates of the Teacher Leaders in Research Based Science Education (TLRBSE; see, e.g., Fitzgerald et al. 2014), a then-ongoing program. TLRBSE was sponsored by the National Science Foundation (NSF), and it touched the formal education community through a national audience of well-trained and supported middle and high school teachers.  In 2010, the Spitzer program was rebranded as NITARP because the source of funding changed to support archival research with teacher teams; applications for participants were then considered from anyone in the US (not just TLRBSE alumni).  In 2013, that funding was steeply curtailed (due to reorganization at NASA).  Between NITARP and its Spitzer predecessor, our model of teacher-scientist partnerships has been refined for 13 years.

The goal of this paper is to describe how NITARP works and share the major lessons learned. This paper begins with an overview of how NITARP works in 2018, and then briefly discusses some of the major 'mileposts' in a NITARP year, with some lessons learned integrate throughout.  Many supporting materials are available on the NITARP website.

**NITARP Overview**

NITARP's year-long program follows the research process, from writing a proposal, collecting data, analyzing data, writing up findings, and then presenting the work at a professional society meeting. Being involved in the whole process can revolutionize teachers' perceptions of "the scientific method'' as it is commonly taught (e.g., Weinburgh 2003). Changing teachers' perceptions of the research process is critical as teachers' perceptions have been shown to impact their pedagogical decisions (Lemberger, Hewson, & Park 1999).

NITARP selects participants from a nation-wide (US) application process. The intent is to engage educators that are not already astronomy experts but have enough scientific background they can come up to speed in a research program quickly.

The program runs from January to January. The "NITARP year" kicks off with a "NITARP Bootcamp" on the day preceding the American Astronomical Society (AAS) meeting, usually during the first full week in January. NITARP pays reasonable travel expenses for the educators to attend the Bootcamp and at least 2 days of the subsequent 4-day AAS meeting. During the Bootcamp, NITARP educators meet their team for the first time. Half of the Bootcamp is reviewing NITARP and the expectations for educators, and the other half of the time is spent in teams, getting to know each other and the science they will do for their project. The teachers return home and work remotely to write a proposal. The proposal is due in March, and it is peer reviewed, by both scientists and educators. Feedback is provided to the proposal writers, and the proposals must be revised in response. Final proposals are posted to the NITARP website. Teams continue to work remotely on their projects through the spring; each team does something different. The teams come out to Caltech in Pasadena, CA for 4 days in the summer. The program pays for reasonable travel expenses for the educators and up to two students per educator. The purpose of this summer trip is to get intensively into the data reduction and analysis for the project. After the visit, the teams return home and continue to work remotely. Abstracts for the AAS meeting are due in October; each team is responsible for at least one science and one education poster at the AAS meeting. Through the rest of the year, the teams finish their work. They go back to an AAS meeting in January to present their results, again with travel paid for the teacher and up to two students per teacher. Participants present their results in the same AAS sessions as professional astronomers, and they must 'hold their own' in that domain; they are not sequestered in a separate session where people know a priori that they are high school teachers and students. Finally, all educators are asked to conduct at least 12 hours of "sharing" in their community, where that could mean professional development, talks at local/regional/national meetings, etc. Over the lifetime of the program, NITARP teams have contributed more than 120 poster papers to the American Astronomical Society (AAS), and contributed to eight refereed papers in major astronomy journals (Rebull et al., 2015, 2013, 2011; Laher et al., 2012ab; Guieu et al. 2010; Howell et al., 2008, 2006).

Because the money that supports NITARP comes from a program that supports archival research, and because the "I" in "NITARP" stands for IPAC, all teams must use at least some of the data housed at Caltech-IPAC. (Earlier in the program's history, some of the money was explicitly tied to outreach associated with Spitzer and others of IPAC's archives.) Fortunately, IPAC is home to very rich archives; about 10% of all refereed astronomy journal articles involve data that originally came from IRSA, just one of the archives at IPAC (IRSA website). We work to create a community of practice among the NITARP alumni (see, e.g., Rebull et al. 2018b), providing ongoing support and a link to the astronomy research community, including tutorial videos on new data and tools.

**Assembling Participants: Team Size**

Every team consists of a mentor scientist, typically three-to-four new educators, and a teacher who has been through NITARP before, called the 'mentor educator.' Smaller teams usually result in too much work per person. More than about five or at most six people per team has proven to be unwieldy (e.g., difficult to find a time when everyone can meet and increased chances of someone not making a deadline such that the whole team has to wait for that person to catch up).

Currently, there are 2 NITARP teams running per year, meaning that there are ~6 new teachers, 2 mentor teachers, and 2 scientists per year. Increasingly, in recent years, there are alumni teams either working entirely on their own, with non-NITARP scientists, or continuing with NITARP scientists "on the side."

The number of concurrent teams is a function of the number of scientists, the number of teachers, and the available money. Money is the limiting factor. Previously, NITARP has had five concurrent teams (not including alumni teams); two concurrent teams (plus additional alumni work) is now more typical.

**Assembling Participants: Finding the Right Scientists**

Finding the right scientists is critical. Scientists **must** see the partnership between them and the teachers as a partnership of equals (for more on partnerships, see, e.g., Johnson et al. 2013). The scientists need to be patient and communicate well; the teachers are skilled, but have different skill sets than an undergraduate. The scientists will very likely improve their teaching skills as well as learn classroom management techniques from the teachers on their team; they need to respect the skills that the teachers bring to the team.

Each scientist must find a project for his or her team that is complex enough to be challenging, and yet simple enough to produce a science poster in a calendar year by educator and student participants who largely do not know how to program in any language. Scientists also must know how long to let the teachers struggle to accomplish a task ("comfortable frustration level") before stepping in and doing it for them, e.g., by coding something up in Python that does a task faster than the teachers can do it by hand or in Excel. Scientists need to be very responsive on email, and be able to sustain a fluctuating time commitment over 13+ months. Much like work with a summer student project, scientists expect to be co-authors on the AAS posters that result from this work, and expect to lead any journal articles that result.

To this point, all NITARP mentor scientists have had some affiliation with IPAC or NOAO (through the TLRBSE heritage). However, several scientists from other institutions have approached NITARP with a desire to mentor a team and/or start a similar program at their home institution. If there is additional money for additional teams, additional scientists can be located. If the program does expand (to more teams or more fields of science), more formalized training (similar to the 'Bootcamp') is likely to be necessary. Such training would include lessons learned from the mentor scientists who have worked with NITARP for its duration; the most important of these lessons learned to date are incorporated into this article.

**Assembling Participants: The NITARP Educator Application Process**

The NITARP application for educators consists of several open-ended questions designed to probe the educators' background and readiness to do research. All past application questions are available on the NITARP website. Broadly, questions cover education background, experience with student research, ability to participate, ability to share the experience, experience with teamwork/online collaboration, and what they hope to get out of the experience.

The NITARP application is released in May. In the past, the application was released later, and some potential applicants reported that they lost access to their email in the summer months, and thus didn't get the NITARP advertising email until it was too late to apply. The application website opens for applications in early August so that teachers can submit their application before they start school, with the deadline in late September. (This date is effectively set by registration deadlines for the AAS.)

The selection committee reviewing the written applications consists of the mentor astronomers for the forthcoming year and external scientists and educators, including a NITARP alumnus. The panel grades the applications independently first, using the same software used by other telescope time allocation committees at IPAC (Crane 2008). Then, the group convenes and discusses the applications in person or over the phone.

Recently, brief (<15 min) online interviews of the finalists have become part of the interview process. Google+ Hangouts are used as a 'hidden' test of computer skills, because they have to install a browser or a plug-in. In-person conversations make it easier to convey how much work the program is, learn better from applicants about their goals and experiences, and see if they are a good match for the program.

Most likely, educators deciding to apply find it appealing to work with NASA scientists (as opposed to, say, Caltech scientists, even though in this context, they are one and the same); the NASA name-brand recognition helps with educator recruiting. If NITARP becomes a model for other programs in other countries or other science disciplines, the role of a "NASA equivalent" in the naming of the program may be important for educator recruitment. Similarly, astronomy is fortunate among the sciences in that most little kids want to be paleontologists or astronomers/astronauts; many adult members of the public retain this affinity and eagerly absorb astronomy outreach. (Examples: Griffith Observatory in Los Angeles gets 1.5 million visitors per year; an overwhelming 45,000 people attending the two-day JPL open house forced JPL to start free timed ticketing admission that caps the number of people who can attend.) Even though few high schools have formal astronomy programs, the appeal of doing astronomy research draws in educators from physics, Earth science, chemistry, and math. A NITARP-style program in other sciences may need to work harder to appeal to potential applicants.

**Assembling Participants: Selecting the Right Educators**

There are always more NITARP educator applicants than spots. Typically at least 4 times as many people apply as can be supported; there is a demonstrable need for this kind of experience. Especially as the Next Generation Science Standards (NGSS 2013; also see A Framework for K-12 Science Education, NRC 2012) are implemented, requiring more inquiry-based and authentic science classroom lessons, teachers will need to find more and more of these kinds of professional development experiences.

NITARP seeks savvy educators, who are already using data with students, and are skilled with computers. Educators must be ready to jump in to research, with a working knowledge of college-level astronomy. At the same time, they can't have already had research experiences. The culmination of the program involves going to the AAS to present their own research; if educators have already done their own research projects and presented the results in a poster (or oral) presentation at the AAS, then the fractional benefit of NITARP to those educators is likely less than for a teacher who has not had these experiences. Since there are so many applicants for so few spots, the program works first to select teachers who have the potential to gain the most from NITARP.

As discussed in more detail in Rebull et al. (2018a), the range of educators who apply to NITARP include the under- and over-qualified. An example of the former would be someone who answers the question about how they involve their students in scientific research by describing how they send their students to the library. An example of the latter would be someone who already has a PhD in astronomy or another physical science; in these cases, at least the institution that granted them their PhD believes that they already understand how scientific research is conducted, so the fractional benefit that NITARP could give them is likely smaller than for other educators. Rebull et al. (2018a) also describes "experience collectors" – these applicants appear to love to add another NASA program to their resume, but don't necessarily put in enough work to be a success in NITARP. The recently implemented online interviews, even as brief as they are, have been of tremendous help in identifying educators as ideal for NITARP or falling into one of the other categories.

Like the mentor astronomers, teachers need to be able to handle lots of email, as well as a fluctuating time commitment, over 13+ months, for free; they also must attend all 3 of the trips associated with the program. NITARP educators must also be US-based; every year, there are inquiries from non-US-based educators, so there is demand even in other countries.

Every year, there are more than enough educators who could benefit from NITARP, and more than enough who are ideally qualified; selections then have to be based not only on the applicant's readiness and suitability for the program, but (as described in the application) also their ability to work on a team, communicate frequently over email, attend all the trips, and share their experience widely and creatively, reaching people that the NITARP scientists would or could not reach on their own.

Selected NITARP educators span a wide range of schools – urban/rural, rich/poor, big/small, private/public, etc. Educators from 34 states have participated, 57% of whom are women. Most of the educators have been public (~65%) high school (~70%) classroom educators (Rebull et al. 2018a). Some middle school teachers have participated, as well as community college educators (those not having advanced degrees in science), and educators from museums or other informal settings.

**Assembling Participants: Finding Mentor Educators**

The mentor educator has been through the program before, and thus supports the scientist in leading the team. He or she helps in translating the scientist to teachers (and vice versa); mentor teachers are good at recognizing confusion among the new teachers and helping the teachers feel comfortable enough to stop the scientist for clarification before proceeding. He or she also helps with logistics, especially as it pertains to navigating school bureaucracy for the trips.

Mentor educator applications are solicited from among the alumni. There is tremendous interest from the alumni community; as a result, there is a cap of three years on the number of times anyone can serve as a mentor educator. The mentor educators also rotate between mentor astronomers so that they can learn different material. Unsurprisingly, the mentor educators report that NITARP scientists approach projects completely differently; they often list that as a significant thing they learned from NITARP.

**The First Trip: Travel Logistics**

The trips are the most exciting part of the NITARP experience for both teachers and students, but the travel logistics can be challenging. Teachers do not often travel for business, let alone on federal funds. For the two later trips, they are invited to bring along their students; traveling with children other than their own can be stressful. Government travel rules require some outlay of cash, which is later reimbursed; this causes anxiety for teachers who may be living paycheck to paycheck.

In response to recommendations from participants, for each trip, NITARP issues a "Big Travel Document", which includes all rules, recommendations, deadlines, examples of what not to do, etc. in one place. This ameliorates some of the travel-related stress.

**The First Trip: NITARP Bootcamp**

The NITARP Bootcamp before the AAS is critical for making sure everyone is on the same page. About half the day is spent talking about NITARP in general terms, and the other half is spent working in the new teams. This workshop necessarily then includes some presentations, but also some group conversations and goal setting, in addition to team bonding.

On the day that the new NITARP class is in the Bootcamp, the teachers in the NITARP class that is finishing up (as well as any self-funded alumni) are traveling to the AAS meeting. By the end of the Bootcamp day, they are likely in town. Anyone affiliated with NITARP who is presenting at the AAS meeting is invited to come to the end of the Bootcamp to share their poster presentation in 3 minutes. This provides practice for the teams who are about to present, and demonstrates the posters to the new class, proving that it is possible to do what they are about to start.

Feeling stupid is part of a scientist's job, and this is so ingrained for most scientists that they no longer notice it (Schwarz 2008). For teachers on these NITARP teams, this is an unfamiliar feeling. Most of them are used to literally being the smartest one in the room (their classroom), and in NITARP they rarely have complete command of all the relevant information, skills, etc. Most of the teachers love this feeling, or hate it, but live with it. However, some educators completely shut down and disengage because it is overwhelming and uncomfortable. Learning from past teachers who disengage, this is now discussed explicitly and often, how it is legitimate to feel stupid and legitimate to not like it, and how this is part of science. We share Schwarz (2008), which is entitled, "The Importance of Stupidity in Scientific Research." Fewer teachers disengage now since this discussion has been implemented. The program reminds participants often through the year that they will not understand everything, and that's ok, and that it's ok to ask questions again and again until they understand.

Additionally, because these teachers do not have research experience, they were most likely taught in college classrooms where "final form science" was emphasized (Duschl 1990). In such science classes, the material was taught as facts—the products of inquiry, but not the scientific inquiry. Occasionally, teachers will ask the research scientist what is the answer they are looking for (e.g., "how many stars are we supposed to find in this dust cloud?"). In NITARP, the teachers and their students are discovering these answers. That makes this experience extremely powerful; in fact, many teachers have described this experience as life-changing (Rebull et al. 2018b).

Another interesting NITARP outcome is that many teachers comment they feel more comfortable talking with scientists or speaking the language of science because of this experience (Rebull et al. 2018b).

Teachers also report that they initially get frustrated with the iterative nature of scientific research. Many teachers arrive at NITARP with the misconception that science is a linear process; one simply cranks through a series of steps to achieve a desired result or gain a bit of knowledge.

While the NITARP teams collaborate well, some issues and conflicts do arise. The most common reasons teams break is lack of regular, open communication. If someone isn't pulling his/her weight at any point during the year, the team will wait for him/her ...but only for a while. Reintegration is impossible after trust is broken in such a fashion, unless information is actually conveyed (e.g., "I didn't complete this month's assignments because my union is on strike" or "My son has been in the hospital."). Outright communication failures hurt ("I missed that email"), but using such failures consistently ("my email is down again") as an excuse does not endear members to the rest of their team. Asking questions often is critical as well. If someone is too confused to pull their weight, they need to ask questions, sooner rather than later, later rather than never. If not, the team breaks.

**The First Trip: AAS Meeting Itself**

AAS meetings can be overwhelming. NITARP provides a worksheet ('treasure hunt') to help give structure to the meeting. It talks about the major reasons people attend the meeting: networking; learning about new science results from presentations both inside and outside of one's field; learning about policy decisions from the relevant federal agencies; visiting booths from observatories, industry, and publishers; learning about educational resources and recent research in astronomy education; and finding posters of collaborators and friends. The worksheet also suggests specific tasks (e.g., 'find the ugliest poster', or 'find someone at an industry booth that doesn't have a PhD'). New participants are reminded that they will be presenting in a year, so they need to identify qualities of effective posters.

**The Teams in "Ordinary Time": Communication During the Year**

In 2005-2006, online collaboration services (such as Google Drive) did not exist. As some became available, schools blocked them, or teachers were expressly forbidden from being on the same service as students. NITARP (then the Spitzer program) started a wiki specifically so as to have a guaranteed option. Recently, schools have been more willing to allow access to these services (many have become 'Google schools'); the use of the wiki has fallen off, though there are still some instances of schools blocking access to some services.

School email systems break often (e.g., mail is not delivered, attachments are stripped or blocked, etc.). Email is a primary communication vector for this program, and having reliable email that is read frequently is critical. To solve this, many teachers have already migrated to gmail or similar services. Interestingly, while most students have email access, they prefer texts or Facebook messages. Educators convey relevant information to their students; one would email her students and then text them to check their email.

Regular group telecons are essential. Teachers do not often work in real time across time zones, though many astronomers do so routinely. It is sometimes difficult to find a schedule for all team members to talk at the same time. If the team does not meet every week or two, the team is often dysfunctional and has trouble making deadlines. For this reason, geography is taken into account when assembling teams, and educators from Maine are not placed on the same team as those from Hawaii. (This is also the primary reason why applications from US educators working at military bases in the rest of the world have had to be turned down; there is just no time that the teams can regularly meet.) Telecons need to be a "safe space" for "dumb" questions (or those feared to be dumb). These meetings are meant primarily to link the scientists with the teachers. Sometimes students listen to the telecon, but some teachers are not comfortable with students on the call; each team handles this decision separately. In recent years, teams have recorded the weekly telecons, posting the recording privately. Recordings provide a way for a teacher who missed a meeting to get caught up, or for them to revisit parts that seemed unclear after the telecon, and/or they can share entire telecons or parts thereof with their students.

Educators have recently suggested that one telecon per month be teachers only so that the teachers can discuss logistics (picking students, negotiating with school administration), or (in some cases) lower the barrier further for asking questions perceived as dumb. Some teams meet occasionally at a different day/time (separate from their call with their scientist), thereby preserving the regular meeting time for science-related questions.

Feedback forms are collected by NITARP at four milestones during the year: before the first AAS, after the first AAS, after the summer visit, and after the second AAS. Using these, in addition to regular group communication, mentor scientists can make sure that all of the educators are keeping up. Changes to how the program runs can be (and are) made during the year in response to these feedback forms.

**The Teams in "Ordinary Time": Project Content**

Each NITARP team does something different, because they are doing new scientific research. However, when the science mentors repeat from year to year, their teams often (but not always!) do similar science; scientists are very specialized and would not feel comfortable leading a research team in a field in which they have no special expertise. No team repeats exactly what was done in a prior year, though they may continue a project begun in an earlier year. For example, one team from 2016 looked for young stars in a particular region using primarily near- and mid-infrared data; the subsequent team in 2017 used mid- and far-infrared data to look for more young stars in the same region and to explore the properties of the previously-identified young stars using longer wavelengths.

Abstracts, project descriptions, and final poster presentations from each team are archived on the NITARP website, so individual specific projects can be explored there (though records are incomplete for the earliest years). Projects have ranged from dusty disks around relatively nearby stars to galaxies at the edge of the Universe. Projects have used data from Earth-orbiting satellites (Widefield Infrared Survey Explorer, WISE; Wright et al. 2010) out to satellites at the Earth-Sun L2 point (Herschel Space Observatory; Pilbratt et al. 2010). Teams have used data from the X-rays (wavelengths, $\lambda$, of $\sim$0.001 µm) to the radio ($\lambda \sim$10,000 µm), but most projects focus on the near infrared ($\lambda \sim$2 µm) to the far-infrared ($\lambda \sim$70 µm), because those data form the heart of the data stored at IPAC. Because of the diversity of data used, as well as the diversity of science goals, any two teams may not use data from instruments that detect photons in the same way, use data that are stored in an archive that can be accessed using the same tools, or even use the same software to measure quantitative things in the data.

**The Teams in "Ordinary Time": Technical Support and Software**

Computer issues can be an enormous challenge. Most teachers and students have Windows machines and most astronomers have Mac or Linux machines, so there is a knowledge gap before the team even starts working. Professional astronomy software may not even be supported for Windows machines. Schools often prohibit software installation, or require months-long lead time before installation. One school in the early years refused to install software because it was free and therefore a purchase order could not be generated. As a result, NITARP teams primarily use common programs like Excel, or OS-independent web-based services. Since archives are moving more towards a model of "analysis at the archive" (e.g., Rebull et al. 2016), in the near future, more research-quality tools will be available in OS-independent web-based formats; this will make it easier for NITARP participants as well as professional astronomers to do research.

Some software has been developed on a volunteer basis as part of the Spitzer program in its early years, and it is still being used. The Aperture Photometry Tool (APT) is a tool for performing aperture photometry; see Laher et al. (2012a, 2012b) for more details. This application is also enjoying a life outside of NITARP in Astro 101 classes nationwide (e.g., R. Kron, priv. comm.).

**The Teams in "Ordinary Time": Support at School**

Teachers must seek permission from their principal or functional equivalent before applying to the program, and must provide assurances that they can attend all the trips.

Teachers tell NITARP that, in order for them to support their negotiations with their administration, NITARP should help them get good press (literally and figuratively) at home. NITARP puts out a press release at the AAS announcing the new class that is starting and announcing the results from the class that is finishing up. Educators provide (beforehand) a list of administrative and/or media contacts. Educators in smaller towns often get quite a bit of media coverage as a result. However, actual media coverage is secondary to making sure that the school administration knows that their NITARP educator is special and doing good things.

**The Teams in "Ordinary Time": Involving Students In General**

NITARP relies on the teachers to select their students, or respects their wishes to learn independently from their students before sharing. NITARP trusts the educators to convey relevant information to their students when they are comfortable doing so; most educators start more intensive student work in late spring.

Some alumni teachers have donated the materials they used to select participating students, and those are on the NITARP website – some have applications with essays, some ask students to summarize a journal article relevant to the research, etc. Some educators select from a large applicant pool; others hand-select a few students. Some educators work with large groups (20 or more) at home, and some just work with a few students; some meet in a designated class, after school, at lunch, or on weekends.

Note that NITARP spans two academic years because it runs January to January, so this must be taken into account when selecting students. Most educators select high school juniors in late spring, near the end of their school year. Those students will still be at the school (as seniors) in the fall/winter, and thus communication is easy. Occasionally, educators bring younger or older students. The emotional needs of younger students are more substantial, and teachers need to more aggressively support those younger students. Older students are harder to connect with during the academic year, and those that leave for college midway through NITARP are hard to engage while the team is finishing their project.

**The Teams in "Ordinary Time": Involving Students on the Trips**

Teachers choose whether to work with any number of students at home, and that is a different decision than choosing whether to bring students on the second and third trips. NITARP relies on the teachers to decide whether to bring students or not – whatever they feel most comfortable doing (or that their school mandates). Most educators bring two or more students.

Some educators raise money to bring more students on the second and third trips. Empirically, most teachers who are responsible for more than four students are distracted simply by the burden of keeping track of many people. Since NITARP's primary goal is to reach the teachers, there is a cap of four students for the summer visit in particular. Teachers rarely bring more than four students to the second AAS.

**Spring: Writing the Proposal**

The first step for many science investigations is writing a proposal, so it is also the first major step in a NITARP project. The purpose and process of writing a proposal helps shape participants' thought process and allows them glimpses of the big picture before diving into the details. The educators also can use the proposal for recruitment of student participants, as well as communications with administrators, so they are grateful to have it. The proposal consists of an abstract, science introduction and context (background on subject, specific target(s), how target(s) were selected and why, and what they expect to find), analysis plan, and an education/outreach plan. The work to create the proposal helps get the educators up to speed on the background as well as having a "story arc" that is the going-in plan.

While no one gets their proposal rejected, the proposal is peer-reviewed, both by scientists and NITARP educator alumni. The teams must respond to their comments, and submit a revised proposal. The proposals are posted on the NITARP website typically in April.

**Spring: After the Proposal**

After the proposal is written, teams do different things. Some teams have held weekly "journal clubs" where papers were discussed in detail. Other teams begin working their data; tools for working with data online have improved dramatically in the last 13 years.

Educators who want to more explicitly involve students typically start in the spring, after the proposal is turned in (see sections on students).

**The Second Trip: The Caltech Summer Visit**

The Summer visit to Caltech is very intense. Though a "research trip" like this is very common for astronomers, usually none of the teachers and students has ever done anything like this. (Evidence suggests that few students have "worked for 8 hours and only stopped to eat once!") This trip does not include the beach or Disneyland, though half a day is spent on a JPL tour given by NITARP staff. During the trip, teachers can be under stress because they are learning side-by-side with students; for some teachers, this is energizing, but not for all. These are long days, spent working on difficult things, and the information transfer rate is very high. This trip is when the team really 'gels' because teachers and students and the scientist are working side-by-side towards a common goal; see Rebull et al. (2018ab) for much more information on the importance of team building and the summer visit.

One frequently asked question in the past, particularly from students, is how much money scientists make. The actual number is meaningless, because the cost of living is very different in Los Angeles compared to, say, rural Oklahoma, and because salaries vary between a telescope operator on a mountaintop and a professor at a tier-1 research university in a big city. In asking this question, what the students are actually asking is whether or not one can sustain 'normal' lives while working as a scientist. To address this larger issue, on the first night of the summer visit (the day everyone arrives), the teams have a pizza party at the mentor scientist's house. By inviting them into their homes, the scientists demonstrate empirically that they have houses and cars and spouses and kids and pets – they have normal lives. Since the implementation of these parties, there are no longer questions about salaries. These parties are also the beginning of substantial team building (particularly for the students who have never met anyone from the other teams before that night) on the eve of the start of the hard work!

The summer visit was originally three days long; teachers strongly recommended a fourth day. However, the scientists are completely drained at the end of three days, and did not think they could sustain a fourth day. Now, the fourth day is "training wheels" – time and space for the team to work on their own, away from their scientist but still all in the same room, before they go home to resume the rest of their regular lives. Typically, the team goes back through notes, making sure they understand all the things that were accomplished during the week, finishing tasks that perhaps some finished that others didn't, and making a schedule to get the project done on time. Consultation with the scientist happens at least once during this fourth day, but the scientist is not 'leading the charge' on this day; the mentor educator typically sets the tone for this last day.

At the end of the summer visit, most teams still have work to do. Remote work continues through the fall. Usually, this translates to far more data analysis than was possible to accomplish in the spring, because the team is more up-to-speed on the required tasks.

Poster abstracts are due to the AAS in October. As for professional astronomers, usually the posters are not even partially done before the abstracts are submitted, but the general tenor of the results is known by the abstract deadline.

**Summer/Fall: Education Work**

In addition to their science poster at the AAS, each team must present at least one education poster. This poster is not particularly supposed to be education research, though the last few years has seen more posters move in that direction. The education poster writing itself is meant to prompt internal reflection on the NITARP experience and to begin the process of integrating the experience into their classroom.

**The Third Trip: Returning to the AAS and the NITARP Retrospective**

The teams' return to the AAS is what they have worked towards for a year. Because the final posters presented by the teams are in the poster sessions appropriate for their respective topics, the posters can be up any of the four days during the main AAS meeting. All the posters are made available on the NITARP website as soon as possible. Each team handles preparations for the AAS differently, though all posters are led by educators. Most of the time, students (stand next to and) present the science poster (with a teacher close by), and teachers present the education poster.

There is a "NITARP Retrospective" on the evening of the first full day of the conference. All NITARP-affiliated people at the meeting are invited: the new class, the finishing class plus students, and any self-funded alumni plus students. There is a big group photo (see Figs. 1 and 2 for two examples), and then the group breaks up into smaller discussion groups consisting of mixtures of people (new teachers, finishing teachers, students, and alumni). Each group discusses what worked, what didn't work, and advice for the next year's class. Answers are collected, and a discussion is held. There are many very good suggestions of things to do to improve the program at these meetings; many of the features of the program as it stands and the lessons learned described above emerged from these discussions. The number one piece of advice they give, every year, to the newest NITARP participants is to ASK QUESTIONS, early and often. These newest NITARP participants report that this meeting is often one of the most useful things they attend during the AAS week. (Sometimes advice coming from peer teachers is heeded more than advice from NITARP management.)

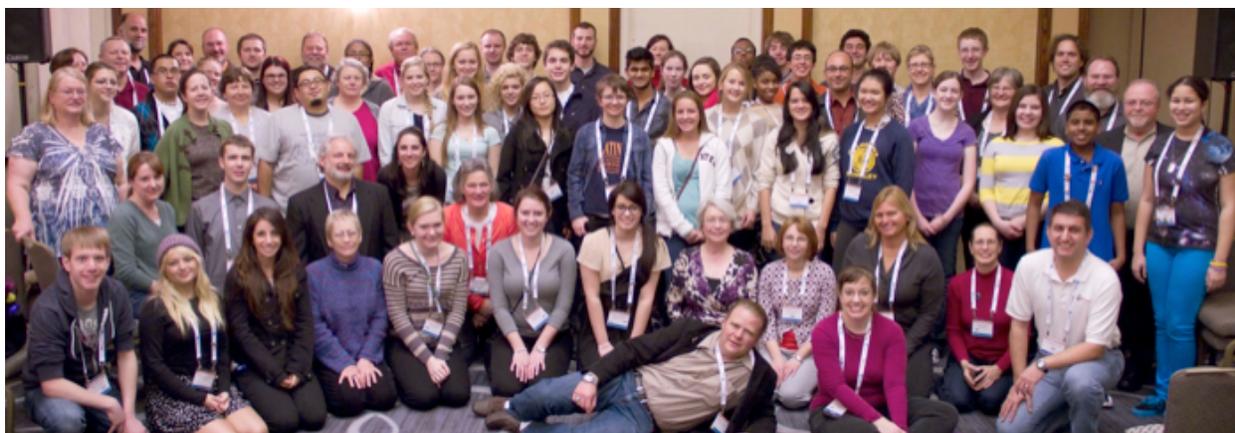
Fig 1. NITARP-affiliated attendees of the 2013 AAS in Long Beach, CA; this represents the 2013 class starting, the 2012 class finishing up (with students) and self-funded alumni (and students). This was the largest ever NITARP delegation; there were about 80 people affiliated with NITARP at this meeting, about 3% of the entire AAS attendance.

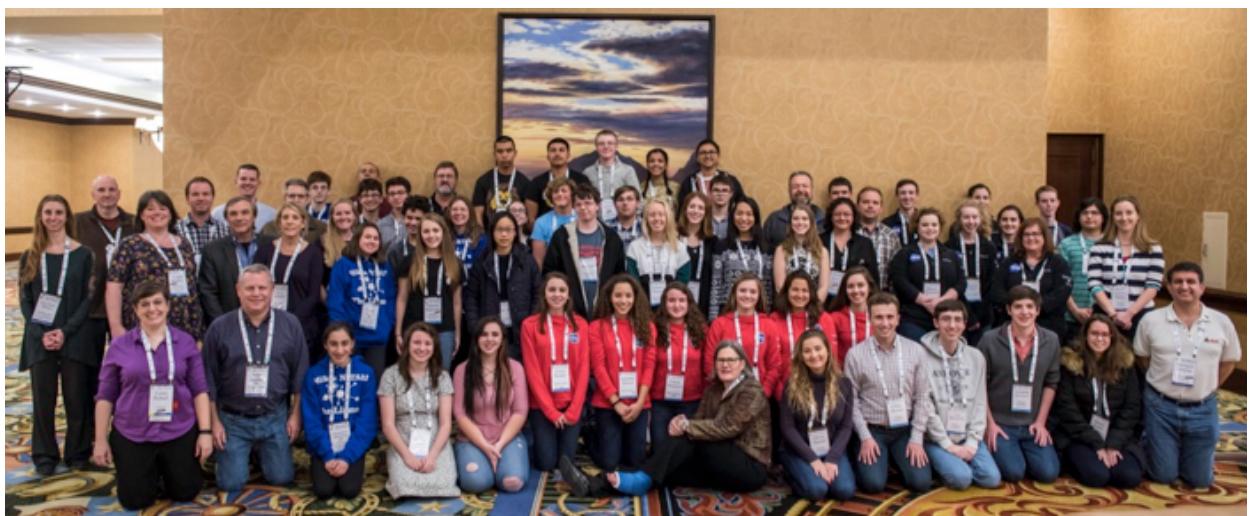
Fig 2. NITARP-affiliated attendees of the 2017 AAS in Grapevine, TX; this represents the 2017 class starting, the 2016 class finishing up (with students) and self-funded alumni (and students). NITARP sent about 50 people, which is more typical of recent delegations. There are about as many alumni educators present in this picture as there are in the 2016 class.

**The Finish Line: Products**

The product of the research is the poster paper (and, for some projects, eventually a journal article).

Teachers involve students on their terms; they do what they need, want, or are able to do. No teacher comes out of NITARP with a solid, ready-to-implement lab or lesson plan or curriculum. Learning side-by-side with their students means that while there is no canned lesson development, the teachers develop real-time lessons for students while learning alongside them. Through their first intensive NITARP year, but also as alumni, they get exposure to resources, and explicitly fold many of these resources into future lessons. Alumni teachers report incorporating programming, authentic data (rather than "canned" data from a textbook), and showcasing the iterative scientific research process (Rebull et al. 2018b). The entire NITARP community is also a long-term resource on which they can (and many do) lean for help in the future (Rebull et al. 2018b).

The educators have an obligation to conduct 12 hours of professional development; they must submit plans for this as part of their application. After the program, they report what they did to share their experience during the subsequent 6-12 months (or more). The teachers are sharing – often repeatedly – but getting them to report what they did is sometimes hard. Most teachers share their experience with other educators through workshops or presentations at all levels – their school and local astronomy clubs, and also district, regional, state, and national meetings (e.g., NSTA, AAPT).

Many alumni have moved up and out of the classroom into higher-level administration or higher education, taking the NITARP experience with them. Tracing this kind of longer-term impact is not something that for which there has yet been resources to study. Some alumni report substantial career changes explicitly as a result of the NITARP experience (Rebull et al. 2018b).

This experience is open-ended by design. Each team may measure 'success' differently. For example, a null result is still valid, and still science, though probably not a journal article – but still a successful NITARP project. A team's work may be a small part of a larger effort being conducted by the scientist, or it might be a small, well-defined project that can be published as a journal article; both of these are successful NITARP projects. Each team studies something different, possibly using vastly different techniques and wavelengths (over many orders of magnitude in wavelength), so the photons are not even collected in the same way, and may not be retrieved from the archives using the same tools, so it is difficult to design, say, a test of core skills for all participants. NITARP has tried a few different approaches for assessment; there was a Summative Evaluation of the 2013 class (Burtnyk 2014). There was a survey of all NITARP participants (then current and alumni) in June 2013 (Rebull et al. 2014). Two more papers look at motivations of educators (Rebull et al. 2018a) and major outcomes of the program (Rebull et al. 2018b); both incorporate more thoughts about what constitutes 'success' in NITARP, and the latter focuses on major changes and outcomes in the educators.

**After NITARP: Alumni community**

The NITARP alumni form an on-going community of practice (Wenger et al. 2002); see Rebull et al. (2018b) for more details. The NITARP mailing list is a place where opportunities are shared and where teachers can ask for help. There is a 'continuing education' video series for NITARP alumni called "NITARP Tutorials," where the videos, created by astronomers, share new tools and data releases with the NITARP community. The videos are posted publically to YouTube, and others learn too; the NITARP Tutorials on FITS viewer ds9 were posted (by the ds9 staff) on Harvard's ds9 page[2].

**Summary**

NITARP, the NASA/IPAC Teacher Archive Research Program, fosters partnerships between teachers and research astronomers. NITARP's goals are to provide a professional development experience for teachers that enables them to experience the real research process, through which their understanding of the nature of research is deepened and ultimately their current and future students are affected via changes in teaching styles. The goal of this paper is to present the structure of the program with many embedded lessons learned, arrived at via more than 10 years of experience, and incorporating many improvements suggested by the participants themselves.

In a calendar year, teams propose a research project, do it, and present the results at an American Astronomical Society winter meeting, which are among the largest astronomy conferences in the world. Participants come from all over the US. Three trips are part of the program (with reasonable expenses paid for by the program): a trip to the AAS to meet the team and get started on learning the science, a trip to Caltech with students to get intensively into the data, and a trip back to the AAS with students to present results. All teams must present at least two posters at the AAS: one science and one education; these posters are presented in the sessions appropriate for their topics (not a special NITARP session).

Scientists need to see the partnership between them and the teachers as a partnership of equals. NITARP educators are selected such that they are already using data with students and are ready to jump into research without having yet done it. Teams consist of a mentor astronomer, a mentor educator, and 3 or 4 new educators; teams much larger or smaller have struggled. Mentor educators have been through the program before and help lead the team. The team must communicate often and honestly; learning new information and skills as fast as necessary in NITARP can be overwhelming and uncomfortable for educators, but most persevere. Teachers involve students on their terms and timescale; a NITARP year spans two academic years, so student involvement does as well. Students frequently ask about scientists' salaries; what they are really asking is whether people can sustain 'normal' lives as scientists. A pizza party at the scientist mentor's house demonstrably (rather than explicitly) answers those questions.

Feedback is collected from participants via surveys at four points during the year, in addition to a large group meeting (the NITARP Retrospective) during the AAS. The program is continuously refined in response to those suggestions; most of the best practices and lessons learned described in this paper have emerged from this process. The product of each team's research is the poster paper (and, for a few projects, a journal article). No ready-to-implement lesson plans are produced; information and resources are incorporated by the educators into their classrooms. A community of practice among the NITARP alumni is maintained for long-term support of the NITARP alumni community.

---

[2] http://ds9.si.edu/site/Documentation.html

Many teachers have only experienced cookbook-style labs and final form science (Crawford, 2014; Duschl, 1990). NITARP offers participating teachers an opportunity to participate authentically in the research process. By doing so, these teachers see there are other ways of teaching and learning about science. Participating in scientific research has been shown to positively impact educators' content knowledge and use of scientific tools and techniques (e.g., Dresner & Worley 2006; Westerlund et al. 2002); it also can have positive effects on their students, even those not involved in the research (e.g., Silverstein et al. 2009). Rebull et al. (2018b) shows that NITARP can be life-changing for participants. The NITARP model is successful and can be expanded (money is the limiting factor, as we have more teachers and astronomers than we can support); it can also be replicated in other sciences.